\documentclass[prd,aps,floats,preprint,preprintnumbers]{revtex4}
\usepackage{amsmath,amssymb}
\usepackage{graphicx}
\usepackage{graphics}
\usepackage{epsfig}
\usepackage{verbatim}
\usepackage{html,makeidx}
\usepackage{color}
\makeindex
\numberwithin{equation}{section}

\newcommand {\sr}{\stackrel}

\newcommand {\eqv}{\equiv}

\newcommand {\ora}{\overrightarrow}
\newcommand {\ola}{\overleftarrow}

\newcommand {\ra}{\rightarrow}
\newcommand {\Ra}{\Rightarrow}

\newcommand {\ral}{\longrightarrow}

\newcommand {\vphi}{\varphi}
\newcommand {\del}{\partial}

\newcommand {\bfr}{\begin{flushright}}
\newcommand {\efr}{\end{flushright}}
\newcommand {\bfl}{\begin{flushleft}}
\newcommand {\efl}{\end{flushleft}}

\newcommand {\nn} {\nonumber}

\newcommand {\txt}{\textrm}
\newcommand {\bd}{
\begin{document}}
\newcommand {\ed}{\end{document}}

\newcommand {\be}{\begin{equation}}
\newcommand {\ee}{\end{equation}}
\newcommand {\bea}{\begin{eqnarray}}
\newcommand {\eea}{\end{eqnarray}}
\newcommand {\ba}{\begin{array}}
\newcommand {\ea}{\end{array}}

\newcommand{\bbib}{\begin{thebibliography}{99}}
\newcommand{\ebib}{\end{thebibliography}}
\newcommand {\bab}{\begin{abstract}}
\newcommand {\eab}{\end{abstract}}

\newcommand {\bc}{\begin{center}}
\newcommand {\ec}{\end{center}}
\newcommand {\bit}{\begin{itemize}}
\newcommand {\eit}{\end{itemize}}
\newcommand {\ul}{\underline}
\newcommand {\txtc}{\textcolor}

\newcommand {\ad}{\textrm{ad}}
\newcommand {\Ad}{\textrm{Ad}}
\def\A{{\cal A}}\def\B{{\cal B}}\def\C{{\cal C}}\def\D{{\cal D}}\def\E{{\cal E}}\def\F{{\cal F}}\def\G{{\cal G}}\def\H{{\cal H}}\def\I{{\cal I}}
\def\J{{\cal J}}\def\K{{\cal K}}\def\L{{\cal L}}\def\M{{\cal M}}\def\N{{\cal N}}\def\O{{\cal O}}\def\P{{\cal P}}\def\Q{{\cal Q}}\def\R{{\cal R}}
\def\S{{\cal S}}\def\T{{\cal T}}\def\U{{\cal U}}\def\V{{\cal V}}\def\W{{\cal W}}\def\X{{\cal X}}\def\Y{{\cal Y}}\def\Z{{\cal Z}}

\def\xbar{\bar{x}}\def\ybar{\bar{y}}\def\zbar{\bar{z}}\def\kbar{\bar{k}}\def\pbar{\bar{p}}


\bd \small

\small
\preprint{SU-4252-885 \vspace{1cm}} \setlength{\unitlength}{1mm}
\title{ Finite Temperature Field Theory on the Moyal Plane
\vspace{0.5cm}}
\author{E. Akofor$^{a}$}\thanks{eakofor@phy.syr.edu}\author{ A. P.
Balachandran$^{a, b, c}$}\thanks{bal@phy.syr.edu}\affiliation{$^{a}$~Department of Physics, Syracuse University, Syracuse, NY
13244-1130, USA\\ $^{b}$~Departmento de Matem$\acute{\txt{a}}$ticas, Universedad Carlos III de Madrid, 28911 Legan$\acute{\txt{e}}$s, Madrid, Spain\\
$^{c}$ C$\acute{\txt{a}}$tedra de Excelencia.}

\begin{abstract}
\vspace{0.5cm}
In this paper, we initiate the study of finite temperature quantum field theories (QFT's) on the Moyal plane. Such theories violate causality which influences the properties of these theories. In particular, causality influences the fluctuation-dissipation theorem: as we show, a disturbance in a space-time region $M_1$ creates a response in a space-time region $M_2$ space-like with respect to $M_1$ ($M_1\times M_2$). The relativistic Kubo formula with and without noncommutativity is discussed in detail, and the modified properties of relaxation time and the dependence of mean square fluctuations on time are derived. In particular, the Sinha-Sorkin result \cite{sorkin-sinha} on the logarithmic time dependence of the mean square fluctuations is discussed in our context.

We derive an exact formula for the noncommutative susceptibility in terms of the susceptibility for the corresponding commutative case. It shows that noncommutative corrections in the four-momentum space have remarkable periodicity properties as a function of the four-momentum $k$. They have direction dependence as well and vanish for certain directions of the spatial momentum. These are striking observable signals for noncommutativity.

The Lehmann representation is also generalized to any value of the noncommutativity parameter $\theta^{\mu\nu}$ and finite temperatures.

\end{abstract}
\maketitle

\newpage
\tableofcontents
\newpage

\section{INTRODUCTION}\label{sec:intro}

The Moyal plane is the algebra $\A_\theta(\mathbb{R}^d)$ of functions on $\mathbb{R}^d$ with the $\ast$-product given by
\bea
\label{moyal1}&&(f\ast g)(x)=f(x)e^{{i\over 2}\ola{\del}_\mu\theta^{\mu\nu}\ora{\del}_\nu}g(x)\eqv f(x)e^{{i\over 2}\ola{\del}\wedge\ora{\del}}g(x),~~f,g\in \A_\theta(\mathbb{R}^d),\\
&&\theta_{\mu\nu}=-\theta_{\nu\mu}=\txt{constant}.
\eea
If $\hat{x}_\mu$ are coordinate functions, $\hat{x}_\mu(x)=x_\mu$, then (\ref{moyal1}) implies that
\bea
[\hat{x}_\mu,\hat{x}_\nu]=i\theta_{\mu\nu}.
\eea
Thus $\A_\theta(\mathbb{R}^d)$ is a deformation of $\A_0(\mathbb{R}^d)$ \cite{qft-us}.

There is an action of a Poincar$\acute{\txt{e}}$-Hopf algebra with a "twisted" coproduct on $\A_\theta(\mathbb{R}^d)$. Its physical implication is that QFT's can be formulated on $\A_\theta(\mathbb{R}^d)$ compatibly with the Poincar$\acute{\txt{e}}$ invariance of Wightman functions \cite{drinfeld,qft-us}. There is also a map of untwisted to twisted fields corresponding to $\theta_{\mu\nu}=0$ and $\theta_{\mu\nu}\neq 0$ (``the dressing transformation''  \cite{grosse,Zamolodchikov}). For matter fields, if these are $\vphi_0$ and $\vphi_\theta$,

\bea
&&\vphi_\theta(x)=\vphi_0(x)e^{{1\over 2}\ola{\del}_\mu\theta^{\mu\nu}P_\nu}\eqv \vphi_0(x)e^{{1\over 2}\ola{\del}\wedge P},\\
&&P_\mu=\txt{Total momentum operator}.
\eea
While there is no twist factor $e^{{1\over 2}\ola{\del}\wedge P}$ for gauge fields, the gauge field interactions of a matter current to a gauge field are twisted as well:
\bea
\H^\theta_I(x)=\H^0_I(x)e^{{1\over 2}\ola{\del}\wedge P},
\eea
where $\H^0_I$ can be the standard interaction $J^{0\mu}A_\mu$ of an untwisted matter current to the untwisted gauge field $A_\mu$.

The twisted fields $\vphi_\theta$ and $\H^\theta_I$ are not causal (local). Thus even if $\vphi_0$ and $\H^0_I$ are causal fields,

\bea
&&[\vphi_0(x),\vphi_0(y)]=0,\\
&&[\H^0_I(x),\H^0_I(y)]=0,\\
&&[\H^0_I(x),\vphi_0(y)]=0,~~x\times y,
\eea
($x\times y$ means that $x$ and $y$ are relatively spacelike),
that is not the case for the corresponding twisted fields. For example,

\bea
&&[\vphi_\theta(x),\H^\theta_I(y)]=e^{-{i\over 2}{\del\over\del x^\mu}\theta^{\mu\nu}{\del\over\del y^\nu}}\vphi_0(x)\H^0_I(y)-e^{-{i\over 2}{\del\over\del y^\mu}\theta^{\mu\nu}{\del\over\del x^\nu}}\H^0_I(y)\vphi_0(x)\neq 0,~~ x\times y.
\eea

Thus acausality leads to correlation between events in space-like regions. The study of these correlations at finite temperatures  at the level of linear response theory (Kubo formula) is the central focus of this paper. We will also formulate the Lehmann representation for relativistic fields at finite temperature for $\theta_{\mu\nu}\neq 0$. It is possible that some of our results for $\theta_{\mu\nu}=0$ and $\theta_{\mu\nu}\neq 0$ are known \cite{fradkin}.

In section 3, we review the standard linear response theory \cite{fradkin} and the striking work of Sinha and Sorkin \cite{sorkin-sinha}.
We also discuss the linear response theory for relativistic QFT's at finite temperature for $\theta_{\mu\nu}=0$. It leads to a natural lower bound on relaxation time, a modification of the result ``$(\Delta r)^2\approx \txt{constant}\times\Delta t$'' of Einstein and its generalization ``$(\Delta r)^2\approx \txt{constant}\times \log\Delta t$'' to the `` quantum regime'' by Sinha and Sorkin \cite{sorkin-sinha}.

Section 4 contains the linear response theory for the twisted QFT's for $\theta_{\mu\nu}\neq 0$. A striking result we find is the existence of correlations between space-like events: A disturbance in a spacetime region $M_2$ evokes a fluctuation in a spacetime region $M_1$ spacelike with respect to $M_2$ $(M_1\times M_2)$. Noncommutative corrections in four-momentum space also have striking periodicity properties and zeros as a function of the four-momentum $k$. They are also direction-dependent and vanish in certain directions of the spatial momentum $\vec{k}$. All these results are discussed in this section.

The results of this section have a bearing on the homogeneity problem in cosmology. It is a problem in causal theories \cite{trodden-vachaspati}. The noncommutative theories are not causal and hence can contribute to its resolution.

In section 5, we derive the finite temperature Lehmann representation for $\theta_{\mu\nu}=0$ and generalize it to $\theta_{\mu\nu}\neq 0$. The Lehmann representation is known to be useful for the investigation of QFT's. The concluding remarks are in section 6.

\section{Review of standard theory: Sinha-Sorkin results}
Let $H_0$ be the Hamiltonian of a system in equilibrium at temperature $T$. It is described by the Gibbs state
\bea
{e^{-\beta H_0}\over \txt{Tr}~e^{-\beta H_0}}
\eea
which gives for the mean value $\omega_\beta(A)$ of an observable $A$,
\bea
&&\omega_\beta(A)={\txt{Tr}~e^{-\beta H_0}A\over \txt{Tr}~e^{-\beta H_0}}.
\eea

We assume that $H_0$ has no explicit time dependence, otherwise it is arbitrary and can describe an interacting system.

We now perturb the system by an interaction $H'(t)$. After this perturbation, the Hamiltonian becomes
\bea
&&H(t)=H_0+H'(t).
\eea

When $H'$ is treated as a perturbation, the change $\omega_\beta(\delta A(t))$ in the expectation value of an observable $A(t)$ in the Heisenberg  picture at time $t$ is
\bea
\omega_\beta(\delta A(t))=\omega_\beta(U^{-1}_I(t)A~U_I(t))-\omega_\beta(A),
\eea
where
\bea
&&U_I(t)=\T e^{-{i\over\hbar}\int_{-\infty}^t d\tau H_I(\tau)}\\
&&H_I(\tau)=e^{{i\over\hbar} H_0\tau}H'(\tau)e^{-{i\over\hbar} H_0\tau}.
\eea
Hence to leading order,
\bea
\omega_\beta(\delta A(t))&=&-{i\over\hbar}\int_{-\infty}^t d\tau~ \omega_\beta([A,H_I(\tau)])\\
         &=&-{i\over\hbar}\int_{-\infty}^\infty d\tau~\theta(t-\tau)\omega_\beta([A,H_I(\tau)]).
\eea
The linear response theory is based on this formula. It is completely general and applies equally well to quantum mechanics and QFT's. But in the latter case, the spatial dependence of the observable should also be specified.

For illustration of known results, we now specialize to quantum mechanics with one degree of freedom and to a dynamical variable $A(t)=x(t)=x(t)^\dagger$ and $H'(t)=x(t)f(t)$ where $f$ is a weak external force. Then,
\bea
&&\omega_\beta(\delta x(t))=-{i\over\hbar}\int^\infty_{-\infty}d\tau~ \theta(t-\tau)\omega_\beta([x(t),x(\tau)])~f(\tau)\\
&&~~~~=\int^\infty_{-\infty}~\chi(t-\tau)~f(\tau),
\eea
where $\chi$ is the susceptibility:
\bea
\chi(t)=-{i\over\hbar}\theta(t)\omega_\beta([x(t),x(0)]).
\eea
Let
\bea
&&W(t)=\omega_\beta(x(t)x(0))=S(t)+iA(t),\\
&&S(t)={1\over2}\omega_\beta(\{x(t),x(0)\}),\\
&&A(t)=-{i\over2}\omega_\beta([x(t),x(0)]).\\
\eea
Then
\bea
\chi(t)={2\over\hbar}~\theta(t)A(t)
\eea

The significant properties of these correlation functions are as follows:
\begin{enumerate}
\item Unitarity:
\bea
\overline{S(t)}=S(t),~~~~\overline{A(t)}=A(t)
\eea
from
\bea
H_0^\dagger=H_0,~~~~x(t)^\dagger=x(t).
\eea

\item Time translation invariance:
\bea
&&S(-t)=S(t)~~~~A(-t)=-A(t)\\
&&~~\Ra~~\overline{W(t)}=W(-t)
\eea
from time independence of $H_0$.

\item The KMS condition: (with $\hbar=1$.)
\bea
W(-t-i\beta)=W(t).
\eea
\end{enumerate}

Denoting the Fourier transform of these functions, including $\chi$, by a tilde ~$\widetilde{}$~, as for instance
\bea
\widetilde{W}(\omega)=\int dt e^{i\omega t}W(t),
\eea
one finds
\bea
&&\widetilde{W}(\omega)=e^{\beta\omega}\widetilde{W}(-\omega),\\
&&\txt{Im}\widetilde{\chi}(\omega)=-{1\over 2}(1-e^{-\beta\omega})\widetilde{W}(\omega),\\
\label{symmetric2corrft}&&\widetilde{S}(\omega)=-\coth{\beta\omega\over 2}\txt{Im}\widetilde{\chi}(\omega).
\eea

The important aspect of these relations is that the dissipative part $\txt{Im}\widetilde{\chi}$ of the (Fourier transform of) susceptibility $\chi$ completely determines all the two point correlations, and hence also the real part $\txt{Re}\widetilde{\chi}$ of $\widetilde{\chi}$.

$\txt{Re}\widetilde{\chi}$ can also be determined from $\txt{Im}\widetilde{\chi}$ the Kramers-Kronig relation \cite{fradkin}.

 Following an argument, presented in \cite{sorkin-sinha}, which exploits the properties of the Heaviside function $\theta$, we can write
\bea
&&\txt{Im}\widetilde{\chi}(\omega)
=-{i\over 2}\widetilde{\chi}'(\omega),\nn\\
\eea
where
\bea
&&\chi'(t):=\txt{sgn}(t)~\chi(|t|),\nn\\
&&\txt{sgn}(t)=\theta(t)-\theta(-t).
\eea
Therefore, (\ref{symmetric2corrft}) becomes
\bea
\label{symmetric3corrft}&&\widetilde{S}(\omega)={i\over 2}\coth{\beta\omega\over 2}~\widetilde{\chi}'(\omega).
\eea
The Fourier transform of (\ref{symmetric3corrft}) gives
\bea
\label{symmetric2corr}S(t)={1\over 2\beta}P\int^\infty_{-\infty}dt'~\txt{sgn}(t'-t)~\chi(|t'-t|)\coth{\pi t'\over\beta},
\eea

where $P$ denotes the principal value of $\coth$.
$\txt{Re}\widetilde{\chi}$ does not contribute to (\ref{symmetric2corr}).

This equation has important physics. In time $\Delta t$, the operator changes by $\Delta x(t)=x(t+\Delta t)-x(t)$. With $t=0$, the square displacement due to equilibrium fluctuations is thus
\bea
\omega_\beta(\Delta x(0)~^2)=2[S(0)-S(\Delta t)]
\eea
so that we obtain the Sinha-Sorkin formula
\bea
&&{1\over 2}\omega_\beta(\Delta x(0)~^2)={i\over 2\beta}P\int _0^\infty dt'\chi(t')[2\coth (\Omega t')-\coth(\Omega(t'+\Delta t))-\coth(\Omega(t'-\Delta t))],\nn\\
&&\Omega={\pi\over\beta}.
\eea
Sinha and Sorkin \cite{sorkin-sinha} have analyzed this equation for the (realistic) ansatz
\bea
\label{ansatz}\chi(t)=\mu[1-e^{-{t\over\tau}}]\theta(t)~\sr{t\gg \tau}{\ral}~ \mu ~\theta(t-\tau),
\eea
where $\tau$ is the relaxation time.

In that case,
\bea
\label{sinha.sorkin.rl}{1\over 2}\omega_\beta(\Delta x(0)~^2)={\mu\hbar\over \pi}\ln {[\sinh(\Omega |\Delta t-\tau|)\sinh(\Omega |\Delta t+\tau|)]^{1\over 2}\over \sinh(\Omega\tau)},
\eea
where we have restored $\hbar$.

Sinha and Sorkin \cite{sorkin-sinha} observed that (\ref{sinha.sorkin.rl}) gives Einstein's relation in the classical regime:
\bea
\label{clregime}\beta\hbar \ll\tau\ll\Delta t: ~~{1\over 2}\omega_\beta(\Delta x(0)~^2)\approx {\mu\over\beta}\Delta t.
\eea
But in addition they found a \emph{logarithmic} dependence of $\Delta t$ in the "quantum" regime:
\bea
\label{qregime}\tau\ll \Delta t\ll \beta\hbar:~~{1\over 2}\omega_\beta(\Delta x(0)~^2)={\mu\hbar\over \pi}\ln{\Delta t\over\tau}.
\eea
They have emphasized that this behavior can be tested experimentally.

They also discuss a regime between the classical and quantum extremes which interpolates (\ref{clregime}) and (\ref{qregime}).

\section{Quantum Fields on Commutative Spacetime}

Hereafter, we set $\hbar=c=1$.

We now specialize to QFT's for $\theta_{\mu\nu}=0$. For simplicity, we take
\bea
H'(t)=e\int d^3y~ N_0(y)\vphi_0(y),
\eea
where $N_0(y)$ is the number density of a charged spinor field $\psi_0$,
\bea
N_0(y)=\psi_0^\dagger(y)\psi_0(y).
\eea
$\vphi_0$ is the externally imposed scalar potential and the subscript denotes that $\theta_{\mu\nu}=0$ for these fields. Again for simplicity, we choose $A$ as well to be the number density at a spacetime point $x$. Then
\bea
&&\omega_\beta(\delta N_0(x))=-{ie\over\hbar}\int d^4y~\theta(x_0-y_0)\omega_\beta([N_0(x),N_0(y)])\vphi_0(y).
\eea
The natural definition of susceptibility in this case is
\bea
\chi_\beta(x,y)=-i{e\over\hbar}\theta(x_0-y_0)\omega_\beta([N_0(x),N_0(y)]).
\eea

With this definition,
\bea
\omega_\beta(\delta N_0(x))=\int d^4y \chi_\beta(x,y)\vphi_0(y).
\eea

We will now analyze this formula.

\begin{flushleft}{\textbf{The Kubo formulae}}\end{flushleft}


The susceptibility $\chi_\beta$ is related to the Wightman function
\bea
W^\beta_0(x,y)={i\over\hbar}\omega_\beta(N_0(x)N_0(y))
\eea
and the autocorrelation and commutator functions
\bea
&&S^\beta_0(x,y)={1\over 2\hbar}\omega_\beta (N_0(x)N_0(y)+N_0(y)N_0(x)),\\
&&A^\beta_0(x,y)={-i\over 2\hbar}\omega_\beta([N_0(x),N_0(y)]),\\
&&\chi_\beta(x,y)=2e\theta(x_0-y_0)A^\beta_0(x,y),\\
&&W^\beta_0(x,y)=S^\beta_0(x,y)+iA^\beta_0(x,y).
\eea

There are more nontrivial conditions coming from the KMS condition which we now discuss.

By assumption, $H_0$ commutes with spacetime translations and rotations as dictated by the Poincar$\acute{\txt{e}}$ algebra. So $\omega_\beta$ enjoys these symmetries and $W^\beta_0(x,y),~S^\beta_0(x,y),~A^\beta_0(x,y)$ depend only on $x_0-y_0$ and $(\vec{x}-\vec{y})^2$. Hence they are even in $\vec{x}-\vec{y}$:
\bea
W^\beta_0(x_0,\vec{x}_0~;~y_0,\vec{y})&=&W^\beta_0(x_0,\vec{y}_0~;~y_0,\vec{x})~~\txt{etc}.\\
 &=& \hat{W}^\beta_0(x_0-y_0~;~(\vec{x}_0-\vec{y})^2).
\eea

As $\hat{W}^\beta_0(x_0-y_0~;~(\vec{x}_0-\vec{y})^2)$ can contain terms with $\theta(x_0-y_0)$, we cannot always claim that it is even in $x_0-y_0$ as well. The same goes for $S^\beta_0$ and $A^\beta_0$.

\subsubsection{Spacelike Disturbances}
If $x$ and $y$ are relatively spacelike,
$[N_0(x),N_0(y)]=0$ because of causality (locality).

So if $\vphi_0=0$ outside the space-time region $D_2$ and we observe the fluctuation in a spacetime region $D_1$ spacelike with respect to $D_2$, then the fluctuation vanishes:
\bea
\omega_\beta(\delta N_0(x))=0~~ \txt{if}~~ x\in D_2,~~ \txt{Supp}\vphi_0=D_2,~~ D_1\times D_2.
\eea
Here Supp denotes the support of the function $\vphi_0$ (it is zero in the complement of the support).

Thus we easily recover the prediction of causality for $\theta_{\mu\nu}=0$ \cite{fradkin}.

\subsubsection{Timelike Disturbances}
In this case, the point of observation $x$ is causally linked to the spacetime region $D_2$. Hence $[N_0(x),N_0(y)]$ need not vanish if $x\in D_1$.

We can model the analysis of this case to the one in Section 2 if $H_0$ is the time translation generator of the Poincar$\acute{\txt{e}}$ group for $\vphi_0=0$. We assume that to be the case.

Following section 2, we now introduce the correlator
\bea
\label{wightmann2fxn}W_0^\beta(x,y)=\omega_\beta(N_0(x)N_0(y)).
\eea
By relativistic invariance, $W_0^\beta$ depends only on $(\vec{x}-\vec{y})^2$. Since $\theta(x_0-y_0)$ is Lorentz invariant when $x-y$ is timelike, it can also depend on $\theta(x_0-y_0)$. Thus $W_0^\beta$ depends on $(\vec{x}-\vec{y})^2$ and $x_0-y_0$ and we can rewrite (\ref{wightmann2fxn}) as
\bea
W_0^\beta((\vec{x}-\vec{y})^2,x_0-y_0)=\omega_\beta(N_0(x)N_0(y)).
\eea
We can thus focus on
\bea
\overline{W}_0^\beta(\vec{x}^2,x_0)=\omega_\beta(N_0(x)N_0(y)).
\eea

It is important that \emph{it is even in} $\vec{x}$.
We cannot say that about $x_0$ because of the potential presence of $\theta(x_0)$.

Now
\bea
&&\overline{W}_0^\beta(\vec{x}^2,x_0)=\omega_\beta(N_0(0)N_0(x))=W_0^\beta(\vec{x}^2,-x_0).
\eea
The presence of $\vec{x}$ thus does not affect the symmetry properties in $x_0$. That is the case also with regard to the KMS condition. We write all these conditions explicitly now: write
\bea
W_0^\beta(\vec{x}^2,x_0)=S_0^\beta(\vec{x}^2,x_0)+iA_0^\beta(\vec{x}^2,x_0),
\eea
where
\bea
&&S_0^\beta(\vec{x}^2,x_0)={1\over 2}\omega_\beta(N_0(x)N_0(0)+N_0(0)N_0(x)),\nn\\
&&A_0^\beta(\vec{x}^2,x_0)=-{i\over 2}\omega_\beta([N_0(x),N_0(0)]).
\eea
Then
\bea
&&\chi_\beta(\vec{x}^2,x_0)=2e\theta(x_0)A_0^\beta(\vec{x}^2,x_0),
\eea
where we have written the susceptibility as a function of $\vec{x}^2$ and $x_0$.
Then as before
\begin{enumerate}
\item $S_0^\beta$ and  $A_0^\beta$ are real functions:
\bea
\overline{S}_0^\beta(\vec{x}^2,x_0)=S_0^\beta(\vec{x}^2,x_0),~~ \overline{A}_0^\beta(\vec{x}^2,x_0)=A_0^\beta(\vec{x}^2,x_0).
\eea
\item $S_0^\beta$ is even in $x_0$ and $A_0^\beta$ is odd in $x_0$:
\bea
{S}_0^\beta(\vec{x}^2,-x_0)=S_0^\beta(\vec{x}^2,x_0),~~ {A}_0^\beta(\vec{x}^2,-x_0)=-A_0^\beta(\vec{x}^2,x_0).
\eea
\item We have the KMS condition
\bea
W_0^\beta(\vec{x}^2,-x_0-i\beta)=W_0^\beta(\vec{x}^2,x_0),
\eea
where we have set the speed of light $c$ equal to $1$.
\end{enumerate}
[We will rewrite ~$\chi_\beta,~\widetilde{\chi}_\beta$~ as ~$\chi^\beta_0,~\widetilde{\chi}^\beta_0$~ to emphasize that they correspond to $\theta_{\mu\nu}=0$.]
Thus from the Fourier transforms distinguished by tildes, as in
\bea
\widetilde{W}_0^\beta(\vec{x}^2,\omega)=\int dx_0~e^{i\omega x_0}W_0^\beta(\vec{x}^2,x_0),
\eea
we get
\bea
&&\widetilde{W}_0^\beta(\vec{x}^2,\omega)=e^{\beta\omega}\widetilde{W}_0^\beta(\vec{x}^2,-\omega),\\
\label{ImSusc1}&&\txt{Im}\widetilde{\chi}^\beta_0(\vec{x}^2,\omega)=-{e\over 2}(1-e^{\beta\omega})\widetilde{W}_0^\beta(\vec{x}^2,-\omega),\\
\label{symmcorr1}&&e\widetilde{S}^\beta_0(\vec{x}^2,\omega)=-\coth{\beta\omega\over 2}\txt{Im}\widetilde{\chi}^\beta_0(\vec{x}^2,\omega)
\eea
 Now following an argument analogous to the one that yielded (\ref{symmetric3corrft}), we are able to write
\bea
\txt{Im}\widetilde{\chi}_0^\beta(\vec{x}^2,\omega)
&=&-{i\over 2}\widetilde{\chi}'{}^\beta_0(\vec{x}^2,\omega),\nn\\
\eea
where
\bea
\chi'{}^\beta_0(\vec{x}^2,x_0)&:=&\txt{sgn}(x_0,\vec{x})~\chi_0^\beta(\vec{x}^2,|x_0|),\nn\\
\txt{sgn}(x_0,\vec{x})&=&\theta(x_0-|\vec{x}|)-\theta(-x_0-|\vec{x}|).\nn\\
\eea
Therefore, (\ref{symmcorr1}) becomes
\bea
\label{symmcorr2}&&e\widetilde{S}^\beta_0(\vec{x}^2,\omega)=-\coth{\beta\omega\over 2}\txt{Im}\widetilde{\chi}^\beta_0(\vec{x}^2,\omega)={i\over 2}\coth{\beta\omega\over 2}\widetilde{\chi}'{}^\beta_0(\vec{x}^2,\omega).
\eea
The Fourier transform of (\ref{symmcorr2}) gives

\bea
e{S}_0^\beta(\vec{x}^2,x_0)={1\over 2\beta}P\int dx'_0~\txt{sgn}(x'_0-x_0,\vec{x})\chi_0^\beta(\vec{x}^2,|x'_0-x_0|)\coth{\pi x'_0\over\beta}.
\eea

The expression for the mean square equilibrium fluctuation $\omega_\beta(\Delta N_0^2)(\vec{x}^2,0)$ follows as before:
\bea
&&{1\over2}\omega_\beta (\Delta N_0^2)((\vec{x}-\vec{y})^2,0)={1\over2}\omega_\beta( (N_0(\vec{x},x_0+\Delta x_0)-N_0(\vec{y},x_0))^2)=e(~S^\beta_0(\vec{0}^2,0)-S^\beta_0((\vec{x}-\vec{y})^2,\Delta x_0)~)\nn\\
&&= {1\over 2\beta}\{~2\int_{|\vec{0}|}^\infty dx'_0~\chi^\beta_0(\vec{0}^2,|x'_0|)~\coth{\pi x'_0\over \beta}\nn\\
&&~~~~-\int_{|\vec{x}-\vec{y}|}^\infty dx'_0~\chi^\beta_0((\vec{x}-\vec{y})^2,|x'_0|)(\coth{\pi(x'_0+\Delta x_0)\over \beta}+\coth{\pi(x'_0-\Delta x_0)\over \beta})~\}\label{fluctuation1}
\eea
So nothing much has changed until this point except for the additional dependence of correlations on $\vec{x}^2$.

An ansatz like (\ref{ansatz}) for susceptibility is no longer appropriate now. That is because if
\bea
x_0^2<\vec{x}^2,
\eea
then as we saw $\chi_0^\beta(\vec{x}^2,x_0)$ is zero by causality.

Thus the relaxation time $\tau$ in units of $c$ has the lower bound $|\vec{x}|$:
\bea
\tau \geqslant |\vec{x}|.
\eea
$\tau$ is a function of $\vec{x}^2$, and we write $\tau(\vec{x}^2)$. Then the generalization of the ansatz (\ref{ansatz}) is
\bea
\chi^\beta_0(\vec{x}^2,x_0)=\mu[1-e^{-{x_0-|\vec{x}|\over\tau(\vec{x}^2)}}]\theta(x_0-|\vec{x}|)~\sr{x_0-|\vec{x}|\gg \tau}{\ral}~ \mu ~\theta(x_0-|\vec{x}|-\tau(\vec{x}^2)).
\eea
This lets us evaluate the mean square fluctuation of number density
\bea
&&\label{m2}{1\over 2}\omega_\beta(\Delta N_0^2)((\vec{x}-\vec{y})^2,0)={\mu\hbar\over \pi} \ln {[\sinh\Omega|\Delta x_0-\tau((\vec{x}-\vec{y})^2)|~\sinh\Omega|\Delta x_0+\tau((\vec{x}-\vec{y})^2)|]^{1\over 2}\over \sinh\Omega \tau(0) },\nn\\
\eea
where ~~$\Omega={\pi\over \hbar\beta}$.

Following Sinha and Sorkin \cite{sorkin-sinha}, we assume that
\bea
\label{bounds}\Delta x_0 \gg \tau(\vec{x}^2)\geqslant |\vec{x}|.
\eea
There are thus four time scales:
\bea
\beta\hbar,~~|\vec{x}|,~~\tau(\vec{x}^2),~~\Delta x_0,
\eea
where we have restored $\hbar$. With the assumption (\ref{bounds}), we have four possibilities to consider:
\begin{enumerate}
\item $\beta\hbar \ll |\vec{x}|\ll \tau(\vec{x}^2) \ll \Delta x_0$,

\item $ |\vec{x}|\ll \beta\hbar\ll \tau(\vec{x}^2) \ll \Delta x_0$,

\item $ |\vec{x}|\ll \tau(\vec{x}^2)\ll  \beta\hbar\ll \Delta x_0$,

\item $ |\vec{x}|\ll \tau(\vec{x}^2)\ll \Delta x_0 \ll \beta\hbar$.

\end{enumerate}

Case 1: \emph{The classical Regime}

Case 1 is the "classical" limit. We get back Einstein's result in this case:

\bea
&&{1\over 2}\omega_\beta(\Delta N_0^2)((\vec{x}-\vec{y})^2,0)\nn\\
&&~~~~={\mu\over \beta}(\Delta x_0-\tau(0))=\mu kT(\Delta x_0-\tau(0)).
\eea

Cases 2 and 3 interpolate the classical regime and the extreme quantum regime of case 4. So let us first consider Case 4.

Case 4: \emph{The Extreme Quantum Regime}

This is the  new regime where Sinha and Sorkin \cite{sorkin-sinha} found a logarithmic dependence on time $\Delta t$ of mean square fluctuations. It is now changed significantly.

\bea
{1\over 2}\omega_\beta(\Delta N_0^2)((\vec{x}-\vec{y})^2,0)={\mu\hbar\over\pi} \ln(~{\Delta x_0\over \tau(0)}[1-({\tau((\vec{x}-\vec{y})^2)\over \Delta x_0})^2]^{1\over 2}~).
\eea

As for the cases 2 and 3, our results are as follows:

\emph{Case 2}: The same as \emph{Case 1}.
\bea
{1\over 2}\omega_\beta(\Delta N_0^2)((\vec{x}-\vec{y})^2,0)={\mu\over \beta}(\Delta x_0-\tau(0)).
\eea

\emph{Case 3}:
\bea
{1\over 2}\omega_\beta(\Delta N_0^2)((\vec{x}-\vec{y})^2,0)={\mu\over\beta}\Delta x_0+{\mu\hbar\over\pi}\ln{\hbar\beta\over 2\pi\tau(0)}.
\eea

\section{Quantum Fields on the Moyal Plane}
 For the Moyal plane, we must use the twisted fields and interactions as explained in the Introduction. That leads to the following expression for $\delta N_\theta$:
 \bea
 \delta N_\theta(x)=-i\int_{-\infty}^\infty dx'_0~\theta(x_0-x'_0)\omega_\beta([N_\theta(x),H_I^\theta(x'_0)]),
 \eea
where
\bea
N_\theta=N_0e^{{1\over 2}\ola{\del}\wedge P},~~H_I(x_0)=e\int d^3 x~\H_I^0(x)e^{{1\over 2}\ola{\del}\wedge P},
\eea
$\H_I^0$ being the interaction Hamiltonian density in the interaction representation.

Note that $e^{{1\over 2}\ola{\del}\wedge P}$ reduces to $e^{{1\over 2}\ola{\del}_0\theta^{0i}P_i}$ on integration over $d^3x$. But we will not use this simplification yet.

We shall first discuss the dependence on $\theta$ of two-point correlators.

Let us first examine the twisted Wightman function:
\bea
&&W_\theta^\beta(x,y)=\omega_\beta(N_\theta(x)N_\theta(y))\nn\\
\label{ncwightmanfxn1}&&~~~~=e^{-{i\over 2}{\del\over\del x^\mu}\theta^{\mu\nu}{\del\over\del y^\nu}}\omega_\beta(N_0(x)N_0(y)e^{-{i\over 2}({\ola{\del}\over\del x^\mu} +{\ola{\del}\over\del y^\mu})\theta^{\mu\nu}P_\nu}).
\eea

We can write this as an integral (and sum) over states with total momentum $p$ such as
\bea
\label{ncwightmanfxn2}\langle p,...|e^{-\beta P_0}N_0(x)N_0(y)e^{-{i\over 2}({\ola{\del}\over\del x^\mu} +{\ola{\del}\over\del y^\mu})\theta^{\mu\nu}P_\nu} |p,...\rangle,
\eea
where the dots indicate that there will in general be many states contributing to a state of given total momentum $p$. We can write (\ref{ncwightmanfxn2}) as
\bea
\langle p,...|e^{-\beta P_0}N_0(x)N_0(y)e^{-{i\over 2}\ola{ad}P_\mu\theta^{\mu\nu}P_\nu} |p,...\rangle,
\eea
where $ad P_\mu A=[P_\mu,A]$. for any operator $A$. But
\bea
\langle p,...|[P_\mu,A] |p,...\rangle=0
\eea
for any $A$. Consequently (\ref{ncwightmanfxn2}) is
\bea
W_\theta^\beta(x,y)=e^{-{i\over 2}{{\del}\over\del x^\mu}\theta^{\mu\nu}{{\del}\over\del y^\nu}}W_0^\beta(x,y).
\eea
But now we can write $W_0^\beta(x,y)$ as we wrote it earlier:
\bea
W_0^\beta(x,y)\ra W_0^\beta((\vec{x}-\vec{y})^2,x_0-y_0).
\eea
It depends on $x-y$. Hence in the exponential,
\bea
{{\del}\over\del x^\mu}\theta^{\mu\nu}{{\del}\over\del y^\nu}=-{{\del}\over\del x^\mu}\theta^{\mu\nu}{{\del}\over\del x^\nu}=0.
\eea
Similarly,
\bea
&&S_\theta^\beta(x,y)={1\over 2}\omega_\theta(N_\theta(x)N_\theta(y)+N_\theta(y)N_\theta(x) )=S_0^\beta((\vec{x}-\vec{y})^2,x_0-y_0),\nn\\
&&A_\theta^\beta(x,y)=-{i\over 2}\omega_\theta([N_\theta(x),N_\theta(y)])=A_0^\beta((\vec{x}-\vec{y})^2,x_0-y_0)
\eea
and they have the properties listed earlier.

But we cannot conclude that $\delta N_\theta$ is independent of $\theta^{\mu\nu}$ as well. Specializing to
\bea
\H_I^0=N_0\vphi_0,
\eea
we find
\bea
&&\delta N_\theta(x)=\delta N_\theta{}^1(x)-\delta N_\theta{}^2(x),\\
&&\delta N_\theta{}^1(x)=-i\int d^4x'~\theta(x_0-x_0')e^{-{i\over 2}{{\del}\over\del x^\mu}\theta^{\mu\nu}{{\del}\over\del x'{}^\nu}}\omega_\beta(N_0(x)\H_I^0(x') e^{-{i\over 2}({\ola{\del}\over\del x^\mu}+{\ola{\del}\over\del x'{}^\mu})\theta^{\mu\nu}P_\nu})
\eea
with a similar expression for $\delta N_\theta^2(x)$.
The last exponential can be replaced by 1 as before. Also, integration over $\vec{x}'$ reduces
~$e^{-{i\over 2}{{\del}\over\del x^\mu}\theta^{\mu\nu}{{\del}\over\del x'{}^\nu}}$~ to ~$e^{-{i\over 2}{{\del}\over\del x^i}\theta^{io}{{\del}\over\del x'{}^0}},$
\bea
e^{-{i\over 2}{{\del}\over\del x^\mu}\theta^{\mu\nu}{{\del}\over\del x'{}^\nu}} \ra e^{-{i\over 2}{{\del}\over\del x^i}\theta^{io}{{\del}\over\del x'{}^0}}.
\eea
Thus
\bea
\delta N_\theta^1=-i\int d^4x'~\theta(x_0-x_0')e^{-{i\over 2}{{\del}\over\del x^i}\theta^{io}{{\del}\over\del x'{}^0}}\omega_\beta(N_0(x)N_0(x'))\vphi_0(x')
\eea
and similarly
\bea
\delta N_\theta^2=-i\int d^4x'~\theta(x_0-x_0')e^{{i\over 2}{{\del}\over\del x^i}\theta^{io}{{\del}\over\del x'{}^0}}\omega_\beta(N_0(x')N_0(x))\vphi_0(x').
\eea

We now discuss the two cases where $x$ is space- and time-like with respect to supp $\vphi_0$.

\begin{flushleft}\emph{$x$ spacelike with respect to Supp $\vphi_0$}:\end{flushleft}

This is the case where we anticipate qualitatively new results.

 While calculating $\delta N^1_\theta(x')-\delta N^2_\theta(x'),$ we cannot set

 \bea
 N_0(x)N_0(x')=N_0(x')N_0(x)~~~~  (\txt{from causality})
 \eea
because the exponentials in the integrand translate the arguments $x$ and $x'$, and can bring them to timelike separations. With this in mind, we can write
\bea
&&\delta N_\theta(x)=-i\int d^4x'~\theta(x_0-x'_0)\cos[{1\over 2}{\del\over\del x^i}\theta^{i0}{\del\over\del x^0{}'}] \omega_{\beta}([N_0(x),N_0(x')])\vphi_0(x')\\
&&~~-\int d^4x'~\theta(x_0-x'_0)\sin[{1\over 2}{\del\over\del x^i}\theta^{i0}{\del\over\del x^0{}'}] \omega_{\beta}(N_0(x)N_0(x')+N_0(x')N_0(x))\vphi_0(x').
\eea
We can replace~ $\cos({1\over 2}{\del\over\del x^i}\theta^{i0}{\del\over\del x^0{}'})$~ by ~$\cos({1\over 2}{\del\over\del x^i}\theta^{i0}{\del\over\del x^0{}'})-1=2\sin^2({1\over 4}{\del\over\del x^i}\theta^{i0}{\del\over\del x^0{}'})$~ as the extra term contributes $0$ by causality. This shows that this term is $O((\theta^{i0})^2)$.
Finally,
\bea
&&\delta N_\theta(x)=-\int d^4x'~\theta(x_0-x'_0)\sin[{1\over 2}{\del\over\del x^i}\theta^{i0}{\del\over\del x^0{}'}] \omega_{\beta}(N_0(x)N_0(x')+N_0(x')N_0(x))\vphi_0(x') \nn\\
&&~~+2i\int d^4x'~\theta(x_0-x'_0)\sin^2[{1\over 4}{\del\over\del x^i}\theta^{i0}{\del\over\del x^0{}'}] \omega_{\beta}([N_0(x),N_0(x')])\vphi_0(x').
\eea

This shows clearly that there is an acausal fluctuation in $\delta N_\theta(x)$ when $\vphi_0$ (the ``chemical potential'') is fluctuated in a region $D_2$ spacelike with respect to $x$.

But it occurs only when time-space noncommutativity $(\theta^{0i})$ is non-zero.

We will come back to this term after also briefly looking at the case where $x$ is not spacelike with respect to $D_2$.

\begin{flushleft}\emph{$x$ is not spacelike with respect to Supp $\vphi_0$}\end{flushleft}

The only change as compared to the spacelike case is that we must restore the extra term, which contributed $0$ in the spacelike case, but does not do that now.

 We can simplify notation by defining $\Delta N_\theta(x)$ for any $x$ as follows:
 \bea
&&\Delta N_\theta(x)=-\int d^4x'~\theta(x_0-x'_0)\sin[{1\over 2}{\del\over\del x^i}\theta^{i0}{\del\over\del x^0{}'}] \omega_{\beta}(N_0(x)N_0(x')+N_0(x')N_0(x))\vphi_0(x')\nn \\
&&~~+2i\int d^4x'~\theta(x_0-x'_0)\sin^2[{1\over 4}{\del\over\del x^i}\theta^{i0}{\del\over\del x^0{}'}] \omega_{\beta}([N_0(x),N_0(x')])\vphi_0(x').
\eea
Then

a) If $x~\times$~Supp~$\vphi_0$,
\bea
\delta N_\theta(x)=\Delta N_\theta(x).
\eea

b) If $x$ is not spacelike with respect to Supp $\vphi_0$,
\bea
\delta N_\theta(x)=i\int d^4x'~\theta(x_0-x'_0) \omega_{\beta}([N_0(x),N_0(x')])\vphi_0(x') + \Delta N_\theta(x).
\eea

\subsection{An exact expression for susceptibility}

We want to write
\bea
\delta N_\theta(x)=\int d^4x'~\chi_\theta(x,x')\vphi_0(x'),
\eea
where $\chi_\theta$ is the deformed susceptibility.

We will succeed in doing that by deriving an exact expression for the Fourier transform
\bea
\widetilde{\chi}_\theta(k)=\int d^4x~e^{ikx}\chi_\theta(x),~~ kx=k_0x_0-\vec{k}\cdot\vec{x},
\eea
in terms of $\widetilde{\chi}_0(k)$. The corrections to $\widetilde{\chi}_0(k)$ have remarkable zeros and direction dependence which we will soon point out.

We can write
\bea
\delta N_\theta(x)=\delta N_0(x)+\Delta N_\theta(x),
\eea
where
\bea
\delta N_0(x)=\int d^4x'~\chi_0(x-x')\vphi_0(x')
\eea
and
\bea
&& \Delta N_\theta(x)=\Delta N^1_\theta(x)-\Delta N^2_\theta(x),\nn\\
&&\Delta N_\theta^{(1)}(x)=-2\int d^4x'~\theta(x_0-x'_0)\sin({1\over 2}{\del\over\del x^i}\theta^{i0}{\del\over\del x'_0})S^\beta_0(x-x')\vphi_0(x')\nn\\
\label{suscept1}&&~~~~:=\int d^4x'~\chi_\theta^{(1)}(x-x')\vphi_0(x'),\\
&&\Delta N_\theta^{(2)}(x)=-4\int d^4x'~\theta(x_0-x'_0)\sin^2({1\over 4}{\del\over\del x^i}\theta^{i0}{\del\over\del x'_0})A^\beta_0(x-x')\vphi_0(x')\nn\\
\label{suscept2}&&~~~~:=\int d^4x'~\chi_\theta^{(2)}(x-x')\vphi_0(x').\\
\eea
In (\ref{suscept1}) and (\ref{suscept2}), ~${\del\over\del x'_0}=({\del\over\del x'_0})_1+({\del\over\del x'_0})_2, $~ where the first differentiates just $S^\beta_0$ and the second differentiates just $\vphi_0$.

On partially integrating the second derivative, it cancels the first derivative acting on $S^\beta_0$ leaving a derivative ${\del\over\del x'_0}$ acting on $\theta(x_0-x'_0)$. So finally
\bea
\chi^{(1)}_\theta(x)=2S^\beta_0(x)\sin({1\over 2}{\ola{\del}\over\del x^i}\theta^{i0}{\ora{\del}\over\del x_0})\theta(x_0)
\eea
and similarly,
\bea
\chi^{(2)}_\theta(x)=-4A^\beta_0(x)\sin^2({1\over 4}{\ola{\del}\over\del x^i}\theta^{i0}{\ora{\del}\over\del x_0})\theta(x_0).
\eea

Let us Fourier transform these expressions setting
\bea
&&\widetilde{\chi}^{(1)}_\theta(k)=\int d^4x~e^{ikx}\chi^{(1)}_\theta(x),\\
&&\widetilde{\chi}^{(2)}_\theta(k)=\int d^4x~e^{ikx}\chi^{(2)}_\theta(x)
\eea
and similarly for $\widetilde{S}(k),~\widetilde{A}(k)$. Then
\bea
&&\widetilde{\chi}^{(1)}_\theta(k)={1\over\pi}\int dx_0~\theta(x_0)[\int dq_0~e^{i(k_0-q_0)x_0}\sin{k_i\theta^{i0}(k_0-q_0)\over 2}~\widetilde{S}(\vec{k},q_0) ],\\
&&\widetilde{\chi}^{(2)}_\theta(k)=-{2\over\pi}\int dx_0~\theta(x_0)[\int dq_0~e^{i(k_0-q_0)x_0}\sin^2{k_i\theta^{i0}(k_0-q_0)\over 4}~\widetilde{A}(\vec{k},q_0) ].
\eea
Here we can write $\widetilde{S}$ and $\widetilde{A}$ in terms of $\txt{Im}\widetilde{\chi}_0$:
\bea
&&\widetilde{S}(\vec{k},k_0)=-\coth{\beta k_0\over 2}~\txt{Im}\widetilde{\chi}_0(\vec{k},k_0),\\
&&\widetilde{A}(\vec{k},k_0)=i\txt{Im}\widetilde{\chi}_0(\vec{k},k_0).
\eea

Finally for the twisted susceptibility $\chi_\theta'$,
\bea
\chi_\theta=\chi_0+\chi_\theta^{(1)}+\chi_\theta^{(2)},
\eea
where we have \emph{exact} expressions for $\chi_\theta^{(j)}$ in terms of $\txt{Im}\chi_0$.

\subsection{Zeros and Oscillations in $\widetilde{\chi}_\theta^{(j)}$}

A generic $\txt{Im}\widetilde{\chi}_0$ is the superposition of terms with $\delta$-function supports at frequencies $\omega$, that is, of terms
\bea
\label{zerosupp}\delta(k_0-\omega)\txt{Im}\widetilde{\chi}^R_0(\vec{k},\omega)
\eea
($R$ standing for ``reduced'').

We now focus on a single frequency $\omega$, that is, the case where $\txt{Im}\widetilde{\chi}_0(\vec{k},k_0)$ equals (\ref{zerosupp}). Then
\bea
&&\widetilde{\chi}_\theta^{1}(k)=-{i\over\pi}\coth{\beta\omega\over 2}~{1\over k_0-\omega}~\sin{k_i\theta^{i0}(k_0-\omega)\over 2}~\txt{Im}\widetilde{\chi}^R_0(\vec{k},\omega)\\
&&\widetilde{\chi}_\theta^{2}(k)={2\over\pi}~{1\over k_0-\omega}~\sin^2{k_i\theta^{i0}(k_0-\omega)\over 4}~\txt{Im}\widetilde{\chi}^R_0(\vec{k},\omega).
\eea

These corrections have striking zeros and oscillations which would be characteristic signals for noncommutativity. Thus,

a)
\bea
\label{zeros1}\widetilde{\chi}^{(1)}_\theta(k)=\widetilde{\chi}^{(2)}_\theta(k)=0~~~\txt{if}~~~{k_i\theta^{i0}(k_0-\omega)\over 2}=2n\pi,~~n\in \mathbb{Z}.
\eea
$\widetilde{\chi}_\theta^{(1)}$ actually vanishes at all $n\pi$.

b) Regarding the oscillations, they are from the $\sin$ and $\sin^2$ terms. The sine repeats if its argument is changed by
\bea
\label{zeros2}2n\pi
\eea
 while the $\sin^2$ term does so if its argument is changed by
 \bea
 \label{zeros3}n\pi
 \eea
$(n\in \mathbb{Z})$.
These are multiplying backgrounds with no particular oscillatory behavior.

Both $a)$ and $b)$ are charecteristic features of the Moyal Plane and in principle accessible to experiments. We emphasize that that both these effects are direction-dependent.

These features may have applications to the homogeneity problem in cosmology \cite{trodden-vachaspati}.

\section{Finite temperature Lehmann representation}
 The Lehmann representation in QFT expresses the two-point vacuum correlation functions of a fully interacting theory in terms of their free field values.  It is exact and captures the properties emerging from the spectrum of $P_\mu$ and Poincar$\acute{\txt{e}}$ invariance in a useful manner.

  We have seen in Section 4 that all the two-point correlations at finite temperature for $\theta^{\mu\nu}\neq 0$ can be expressed in terms of the corresponding expressions for $\theta^{\mu\nu}= 0$.
  In this section, we treat the $\theta^{\mu\nu}=0$ case in detail which then also covers the $\theta^{\mu\nu}\neq 0$ case.

  First we state some notation. The single particle states are normalized according to
  \bea
  \label{normalzn}\langle k'|k\rangle=2|k_0|\delta^3(k'-k),~~k_0=(\vec{k}^2+m^2)^{1\over 2},
  \eea
  where $m$ is the particle mass. The scalar product of $n$-particle states such as $|k_1,...,k_n\rangle$ then follows, (with appropriate symmetrization factors which we will not display here or below).
  We will also not display degeneracy indices such as those from color: their treatment is easy. For a similar reason, we consider spin $0$ fields.

  For the normalization (\ref{normalzn}), the volume form $dV_n$ for the $n$-particle state is a product of factors

 ${d^3k_j\over 2|k_{j0}|}$:
 \bea
 dV_n=\prod_{j=1}^nd\mu_j,~~d\mu_j={d^3k_j\over 2|k_{0j}|},~~|k_{j0}|=\sqrt{\vec{k}_j^2+m_j^2}.
  \eea

Now consider
\bea
W_0^\beta(x)=\omega_\beta(\vphi_0(x)\vphi_0(x')),
\eea
where $\vphi_0$ is a scalar field for $\theta^{\mu\nu}=0$ and $H$ is the total time-translation generator of the Poincar$\acute{\txt{e}}$ group. Its spacetime translation invariance implies that
\bea
\omega_\beta(\vphi_0(x)\vphi_0(x'))=\omega_\beta(\vphi_0(x-x')\vphi_0(0)).
\eea

We assume as usual that
\bea
\label{vacuumcondn}\langle 0| \vphi_0(x)| 0\rangle=0.
\eea

We can write
\bea
&&W_0^\beta(x)={\langle 0| e^{-\beta H}\vphi_0(x)\vphi_0(0)|0 \rangle +\omega_\beta(\vphi_0(x)|0 \rangle\langle 0|\vphi_0(0))\over Z(\beta)}+\widehat{W}_0^\beta(x),\\
&&Z(\beta):=\txt{Tr}e^{-\beta H}.
\eea

We shall see that the vacuum contributions are separated out in the first two terms and that vacuum intermediate states do not contribute to $\widehat{W}_0^\beta$.

We now consider the three terms separately.

\bea 1)~~~~{1\over Z(\beta)} \langle 0| e^{-\beta H}\vphi_0(x)\vphi_0(0)|0 \rangle={1\over Z(\beta)}W_0^0(x)\eqv {1\over Z(\beta)}W(x).
\eea
Here $W(x)$ is the zero-temperature Wightman function with its standard spectral representation:
\bea
W(x)=\int dM^2~\rho(M^2)\Delta_+(x,M^2),~~\Delta_+(x,M^2)=\int d^4p~\delta(p^2-M^2)\theta(p_0)e^{ipx}.
\eea

\bea
2)~~~~ \omega_\beta(\vphi_0(x)|0\rangle\langle 0|\vphi_0(0))={1\over Z(\beta)}\sum_{n\geqslant 1}\int dV_n~\langle k_1,...,k_n|e^{-\beta H}\vphi_0(x)|0\rangle\langle 0|\vphi_0(0)|k_1,...,k_n\rangle
\eea
where the $n=0$ term has been omitted in the sum as it contributes $0$ by (\ref{vacuumcondn}).

Using
\bea
\vphi_0(x)=e^{iPx}\vphi_0(0)e^{-iPx},
\eea
where $P_\mu$ generates translations $(P_0=H)$, we find
\bea
&&\omega_\beta(\vphi_0(x)|0\rangle\langle 0|\vphi_0(0))={1\over Z(\beta)}\int d^4 k~\theta(k_0)e^{-\beta k_0+ikx}\rho(k^2),\\
&&\rho(k^2)=\sum_n\int \prod_{j=1}^n~\delta(k^2_j-m^2_j)\theta(k_{j0})\delta^4(\sum k_j-k)~|\langle k_1,...,k_n|\vphi_0(0)|0\rangle|^2,
\eea
$\rho$ being the zero-temperature spectral function.

Thus
\bea
&&\omega_\beta(\vphi_0(x)|0\rangle\langle 0|\vphi_0(0))={1\over Z(\beta)}\int dM^2~\rho(M^2)\Delta_+(x,M^2;\beta),\\
&&\Delta_+(x,M^2;\beta)=\int d^4k~\theta(k_0)\delta(k^2-M^2)e^{-\beta k_0+ikx}.
\eea
For $\beta=0$, $\Delta_+(x,M^2;0)$ is the free field zero-temperature Wightman function. It vanishes when $\beta\ra \infty$.

\bea
3)~~~~ \widehat{W}_0^\beta(x)={1\over Z(\beta)}\sum_{n,m\geqslant 1}\int dV_n dV_m~\langle k_1,...,k_n |e^{-\beta H}\vphi_0(x)|q_1,...,q_m\rangle~\langle q_1,...,q_n |\vphi_0(0)|k_1,...,k_m\rangle.\nn\\
\eea
The vacuum contributions ($n$~and /or $m=0$) have already been considered and need not be included here.

Elementary manipulations like those above show that
\bea
&&\widehat{W}_0^\beta(x)={1\over Z(\beta)}\int d^4K d^4Q~\theta(K_0)\theta(Q_0)e^{-\beta K_0+i(K-Q)x}\times\nn\\
&&~~~~\{~\sum_{n,m\geqslant1}\int\prod_{j=1}^nd^4k\theta(k_{j0})\delta(k^2_j-m^2_j)\prod_{j=1}^md^4q\theta(q_{j0})\delta(q^2_j-m^2_j)\times\nn\\
&&~~~~\delta^4(\sum k_j-K)\delta^4(\sum q_j-Q)~|\langle k_1,...,k_n|\vphi_0(0)|q_1,...,q_m\rangle|^2\}.\nn\\
\eea
The term in braces, by relativistic invariance, depends only on $K^2,~Q^2$ and $(K+Q)^2$.
As $K_\mu,~Q_\mu$ are timelike with $K_0,Q_0>0$, we have, as in scattering theory,

\bea
(K+Q)^2\geqslant (\sqrt{K^2}+\sqrt{Q^2})^2.
\eea
Call the terms in braces as $\rho(K^2,Q^2,(K+Q)^2)$. Then

\bea
&&\widehat{W}_0^\beta(x)={1\over Z(\beta)}\int dM^2dN^2dR^2~\rho(M^2,N^2,R^2)\times\nn\\
&&~~~~\{~\int d^4K~\theta(K_0)\delta(K^2-M^2) \int d^4Q~\theta(Q_0)\delta(Q^2-N^2)~\delta((K+M)^2-R^2)e^{-\beta K_0+i(K-Q)x} ~\}.\nn\\
\eea

The term in braces here is the elementary function appropriate for $\widehat{W}_\theta^\beta$.

The full spectral representation for $W_\theta^\beta$ is obtained by adding those of its terms given above.

\section{Conclusions}A major result of this paper is the derivation of acausal and noncommutative effects in finite temperature QFT's. They are new and are expected to have applications for instance in the homogeneity problem in cosmology. We plan to return to this topic elsewhere.

We have also treated the finite temperature Lehmann representation on the commutative and Moyal planes in detail. This representation succintly expresses the spectral and positivity properties of the underlying QFT's in a transparent manner and are thus expected to be useful.

\begin{center}
\textbf{Acknowledgements}
\end{center}
A. P. B. acknowledges V. P. Nair's help with references. This work was supported by the US Department of Energy under grant number DE-FG02-85ER40231 and by the Universidad Carlos III de Madrid. A. P. B. thanks T. R. Govindarajan and Alberto Ibort for their wonderful hospitalities at the Institute of Mathematical Sciences, Chennai and the Universidad Carlos III de Madrid respectively.

\newpage
\bibliographystyle{apsrmp}

 \ed